\documentclass[useAMS,usenatbib]{mn2e}
\usepackage{graphicx}%for EPS, PS files

\footnotesize
\newdimen\digitwidth    %define ! a one digit width for tables
\setbox0=\hbox{\rm0}
\digitwidth=\wd0
\catcode`!=\active
\def!{\kern\digitwidth}
\normalsize

\title[A white dwarf companion to PSR\,J1141$-$6545]{A white dwarf companion to the relativistic pulsar PSR\,J1141$-$6545\thanks{Based on observations made with ESO Telescopes at the Paranal Observatories under programme ID 381.D-0852(A).}}

\author[J. Antoniadis et al.]{J.~Antoniadis,$^1$ \thanks{Member of the International Max Planck Research School (IMPRS) for Astronomy and Astrophysics at the Universities of Bonn and Cologne. e-mail: jantoniadis@mpifr-bonn.mpg.de} C.~G.~Bassa,$^2$ N.~Wex,$^1$ M. Kramer$^1$ and R.~Napiwotzki$^3$
\\
$^{1}$Max-Planck-Institut f\"{u}r Radioastronomie, Auf dem H\"{u}gel 69, 53121 Bonn, Germany
\\
$^{2}$Jodrell Bank Centre for Astrophysics, The University of Manchester, Alan Turing Building, M13 9PL Manchester, United Kingdom
\\
$^{3}$Centre of Astrophysics Research, University of Hertfordshire, College Lane, AL10 9AB Hatfield, United Kingdom}
\let\leqslant=\leq %Authors with AMS fonts and mssymb.tex can comment
\date{}
\begin{document}

\maketitle

\newcommand{\setthebls}{
%                 de-comment this line for double spacing:
 %\baselineskip=20pt
}

%\setthebls
%%%%%%%%%%%%%%%%%%%%%%%%%%%%%%%%%%%%%%%%%%%%%%%%%%%%%%%%%
\begin{abstract}
Pulsars with compact companions in close eccentric orbits are unique laboratories for testing general relativity and alternative theories of gravity. Moreover, they are excellent targets for future gravitational wave experiments like LISA and they are also highly important for understanding the equation of state of super-dense matter and the evolution of massive binaries. Here we report on optical observations of the 1.02\,M$_{\odot}$ companion to the pulsar PSR\,J1141$-$6545. We detect an optical counterpart with apparent magnitudes $V=25.08(11)$ and $R=24.38(14)$, consistent with the timing position of the pulsar. We demonstrate that our results are in agreement with a white dwarf companion. However the latter is redder than expected and the inferred values are not consistent with the theoretical cooling tracks, preventing us from deriving the exact age. Our results confirm the importance of the PSR\,J1141$-$6545 system for gravitational experiments.
\end{abstract}
%%%%%%%%%%%%%%%%%%%%%%%%%%%%%%%%%%%%%%%%%%%%%%%%%%%%%%%%%%
\begin{keywords}
methods: photometry -- binaries: close -- pulsars: general -- stars: neutron -- white dwarfs -- individual: PSR\,J1141$-$6545
\end{keywords}

%%%%%%%%%%%%%%%%%%%%%%%%%%%%%%%%%%%%%%%%%%%%%%%%%%%%%%%%%%

\section{Introduction}
The value of relativistic binaries is highly recognised, as their study can provide insight into some of the holy grails of fundamental physics. Among them are tests of general relativity and alternative theories of gravity, the detection of gravitational waves, the study of the equation of state of super-dense matter and tests of evolutionary scenarios for heavy stars \citep[for a complete review see][]{2004hpa..book.....L}.

The sample of relativistic binaries discovered so far is dominated by double neutron stars, covering a wide range of orbital parameters. Another substantial fraction consists of white dwarf--neutron star binaries, most of them in almost perfectly circular orbits \citep[e.g. review by][]{2005ASPC..328..357V}. These systems are the result of the evolution of a massive primary which evolves fast, explodes as a supernova and becomes a neutron star (NS); and of a lighter secondary which evolves slower and eventually becomes a white dwarf (WD) \citep{1998A&A...339..123D}. During the final interaction phase, the NS is spun up to very short rotation periods and becomes a millisecond pulsar.  Any eccentricity (primordial or resulting from the supernova kick) is dampened by tidal interaction before the secondary becomes a WD.

A significant exception to the preceding is the binary PSR\,B2303+46 \citep{1985ApJ...294L..21S}. In that system, the WD \citep{1999ApJ...516L..25V} orbits the non-recycled pulsar in a highly eccentric orbit. Investigations into possible formation scenarios for this type of binary 
\citep{2000A&A...355..236T,2002MNRAS.335..369D,2006MNRAS.372..715C} have shown that they most likely originate from a binary system of massive stars with nearly equal mass. When the initially more massive star reaches the red giant phase, the secondary star accretes sufficient mass to surpass the Chandrasekhar limit, allowing it to eventually evolve into a NS. The primary star, however, loses sufficient mass to end up as a heavy WD. Hence, in the resulting system, the WD is expected to be older than the pulsar.

The only other promising candidate for this category is the PSR\,J1141$-$6545 binary system, initially discovered in a Parkes survey \citep{2000ApJ...543..321K}. PSR\,J1141$-$6545 is a $0.2$\,day binary in an eccentric orbit \citep[$e\sim0.17$,][]{2008PhRvD..77l4017B}. The primary is a relatively young 394\,ms pulsar (characteristic age $\sim 1.4$\,Myrs), orbited by a compact object of unknown nature. \citet{2008PhRvD..77l4017B} derived $M_{\mathrm c}=1.02(1)$\,M$_{\odot}$ for the mass of the companion by applying the relativistic \rm{DDGR} orbital model \citep{1986AIHS...44..263D}  to their timing measurements. The latter is consistent with both a heavy WD and a light NS with the former case being more favoured by statistical evidence \citep{2000A&A...355..236T}. \citet{2006ApJ...640L.183J} included the system in an optical survey but found no optical counterpart down to $R=23.4$.

This paper reports on optical observations of the companion star in the PSR\,J1141$-$6545 binary system. Our main scientific rationale for this study is that in the case of a positive WD confirmation, the system would be of great importance for gravitational tests. In particular, because of its gravitational asymmetry, PSR\,J1141$-$6545 would be one of the most constraining systems known for general relativity in the strong field regime as it is expected to emit strong dipolar gravitational radiation in a wide range of scalar-tensor theories \citep{1993tegp.book.....W,2005tmgm.meet..647E,2008PhRvD..77l4017B}.

The structure of the text is as follows: in Section~\ref{section:2} we describe the observations and the data reduction process while in Section~\ref{section:3} we present our results. Finally, in Section~\ref{section:4} we discuss our findings and comment on their astrophysical consequences and their importance in gravitational tests.

%%%%%%%%%%%%%%%%%%%%%%%%%%%%%%%%%%%%%%%%%%%%%%%%%%%%%%%%%

\section{Observations and data reduction}\label{section:2}
We have obtained optical images in the $V$-band and $R$-band filters, of the field containing PSR\,J1141$-$6545 using the \rm{FORS1} instrument mounted at the UT2 of the Very Large Telescope (VLT). Both filters resemble the standard Johnson-Cousin filters but have slightly higher sensitivity in the red, sharper cut offs and higher throughput. The observations were conducted in service mode during the night of 6th of April 2008. The conditions were photometric and the average seeing of the night was $0\farcs7$. The total exposure time was 600 seconds in $V$ and 1500 seconds in $R$. In order to minimize potential problems with cosmic rays and guiding errors and avoid saturation of bright stars, the exposures were split in three sub-exposures of 200 seconds in the $V$-band and three sub-exposures of 500 seconds in the $R$-band.
For the data reduction we used the \rm{FORS1} pipeline provided by ESO. Each image was first bias corrected and flat-fielded using twilight flats. Bad pixels and cosmic ray hits in all frames were replaced by a median over their neighbourghs. The resulting frames were then sky-subtracted, registered and combined in one averaged frame for each filter.
%%%%%%%%%%%%%%%%%%%%%%%%%%%%%%
\subsection{Photometry}
We performed point-spread function (PSF) photometry on the average frame of each filter using \rm{DAOPHOT\,II} \citep{1987PASP...99..191S} inside the Munich Image Data Analysis System (\rm{MIDAS}). The PSF was determined following a slightly modified version of the recipe in \citet{1987PASP...99..191S}.
First, we selected 100 bright, unsaturated stars ($\leq 40000$ ADUs) located within $1'$ distance from our target. Then we fitted their PSFs with a Moffat function and through an iterative process we rejected fits with root mean square (rms) residuals greater than 1 per cent. The stars in the vicinity of the PSF template stars were then removed with the \rm{SUBTRACT} routine of \rm{DAOPHOT\,II} and the PSF was determined again on the subtracted image, improving the rms of the fit by a factor of $\sim$2. Finally, the instrumental magnitudes of all stars within the same distance were extracted.

For the photometric calibration we first found the offset between PSF and aperture magnitudes of six isolated bright stars in both our science images. This offset was used to transform the extracted PSF magnitudes to aperture ones. 
Zero-points and colour terms were determined by analysing two archival images of NGC\,2437 (one in each band), obtained during the 5th of April 2008. The latter contains more than 80 Stetson photometric standards \citep{2000PASP..112..925S}. Of those, we used only 30 depicted on the same area of the CCD as our target. We determined their instrumental magnitudes using the same aperture, inner and outer sky radii as in our science images. We fitted for zero-points and colour terms using the average extinction coefficients provided by ESO (0.120(3) and 0.065(4) per airmass for $V$ and $R$ respectively) and used them to transform our measurements to the standard Johnson-Cousin system. The rms residual of the fit was 0.02\,mag in $V$ and 0.04\,mag in $R$.
%%%%%%%%%%%%%%%%%%%%%%%%%%%%%%
\subsection{Astrometry}
   For the astrometric calibration we selected 58 astrometric
   standards from the USNO CCD Astrograph Catalogue \citep[\rm{UCAC3},][]{2010AJ....139.2184Z}
    that coincided with the $7\arcmin\times4\arcmin$
   averaged $V$ image. Because of the 200\,sec exposure times,
   only 13 of them were not saturated or blended and appeared
   stellar. The centroids of these stars were measured and an
   astrometric solution, fitting for zero-point position, scale and
   position angle, was computed. Two outliers were iteratively
   removed, and the final solution, using 11 stars, had
   root-mean-square residuals of $0\farcs056$ in right ascension and
   $0\farcs058$ in declination, which is typical for the \rm{UCAC3}
   catalogue.

   The low number of astrometric standards used makes the astrometric
   solution sensitive to random noise, and hence we computed another
   solution using stars from the \rm{2MASS} catalogue. Of the
   360 stars from this catalogue that coincided with the $V$ image,
   245 were not saturated and appeared stellar and unblended. The
   iterative scheme removed outliers and converged on a solution using
   210 stars with rms residuals of $0\farcs14$ and $0\farcs13$. This
   solution is consistent with the \rm{UCAC3} astrometric solution to
   within the uncertainties and we are confident in using the \rm{UCAC3}
   solution for the astrometric calibration.
 
 %%%%%%%%%%%%%%%%%%%%%%%%%%%%%%%%%%%%%%%%%%%%%%%%%%%%%%%%%
\section{Results} \label{section:3}
\begin{figure}
\resizebox{8.5cm}{!}{\includegraphics{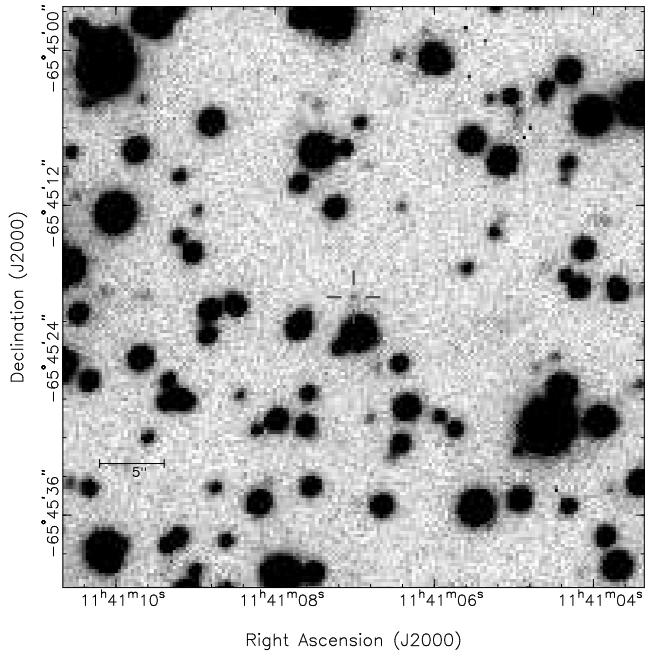}} 
\caption{A $45\arcsec\times45\arcsec$ subsection of the averaged
        V-band image. The timing position of PSR\,J1141$-$6545 is
        denoted by $1\arcsec$ tickmarks.}
\label{figure:1}
\end{figure}

\begin{figure}
\resizebox{8.5cm}{!}{\includegraphics{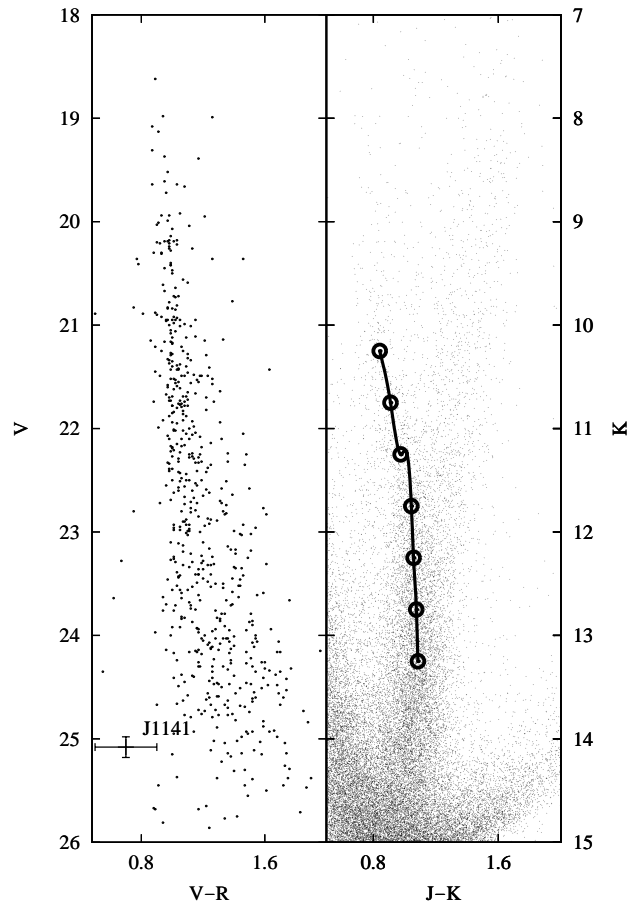}} 
\caption{\textbf{Left}: The extracted $V$ magnitudes of all objects in our data, plotted against their $V-R$ colours. The counterpart of PSR\,J1141$-$6545 is placed among the faintest and bluest objects. \textbf{Right}: Colour$-$Magnitude diagram of 2MASS sources, located within $20\arcmin$ distance from PSR\,J1141$-$6545. The circles indicate the calculated position of red clump stars. The line is a 3$^{\rm rd}$ order spline connecting all circles.}
\label{figure:2}
\end{figure}

A faint star is present on the timing position of the pulsar
  \citep{2010ApJ...710.1694M} in both the averaged $V$ and $R$ images
  (Fig.\,\ref{figure:1}). The optical position is
  $\alpha_\mathrm{2000}=11^\mathrm{h}41^\mathrm{m}07\fs00(2)$ and
  $\delta_\mathrm{2000}=-65\degr45\arcmin19\farcs01(10)$, where the
  uncertainty is the quadratic sum of the positional uncertainty of
  the star (approximately $0\farcs08$ in both coordinates) and the
  uncertainty in the astrometric calibration. This position is offset
  from the timing position by $\Delta\alpha=-0\farcs08\pm0\farcs11$
  in right ascension and $\Delta\delta=0\farcs10\pm0\farcs12$ in declination.
  Hence,  the timing and the optical positions agree within errors.
 The images have an average stellar density of 239 stars per square
  arcminute, which translates to only a 0.9 per cent probability of a chance
  coincidence within the 95 per cent confidence error circle, which has a
  radius of $0\farcs20$. The star has $V=25.08(12)$ and
  $R=24.38(14)$ and at $V-R=0.70(18)$, it is significantly
  bluer than the bulk of the stars in the field, which have
  $V-R=1.27(30)$ for $24<V<26$. Any MS or post-MS star 
  would be brighter and/or redder given the distance 
  ($\geq 3.7$\,kpc, Section~\ref{section:3}), hence we are confident that the
  star inside the error circle is the white dwarf companion to
  PSR\,J1141$-$6545 (Fig.~\ref{figure:2}).
%%%%%%%%%%%%%%%%%%%%%%%%%%%%%%%%%%%%%%%%%%%%%%
\subsection{Distance and reddening}
The intrinsic color and brightess of the WD and hence its cooling age and temperature, can be inferred from our measurements under the condition of an accurate distance and reddening estimate. Unfortunately, as for most pulsars, the distance to the PSR\,J1141$-$6545 system is not well known.

An estimate can be made from the observed dispersion measure (DM) and a model of the free electron distribution in the Galaxy. Using the \rm{NE2001} Galactic free electron model \citep{2002astro.ph..7156C}, we find $d=2.4$\,kpc for the observed $\mathrm{DM}=116.08$\,cm$^{-3}$pc \citep{2010ApJ...710.1694M} towards PSR\,J1141$-$6545. Traditionally the uncertainty on DM derived distances is quoted at 20 per cent, however, a comparison with pulsar parallaxes indicate that the uncertainties may be as large as 60 per cent \citep{2009ApJ...701.1243D}. \citet{2002MNRAS.337..409O} placed a lower bound on the distance by measuring the H{\sc i} absorption spectrum of the pulsar. They concluded that the binary must be located beyond the tangent point predicted by the Galactic rotation model of \citet{1989ApJ...342..272F} to be at $3.7$\,kpc.

The interstellar extinction towards PSR\,J1141$-$6545 was traced using the red clump stars method described in \citet{2006ApJ...650.1070D}. We used a sample of 44168 stars from the 2MASS catalogue, situated within $20\arcmin$ distance from PSR\,J1141$-$6545 (right panel of Fig.~\ref{figure:2}). We then split the sample in seven 0.5\,mag$-$wide stripes, ranging from $K=10$ to $K=13.5$ and traced the $J-K$ location of the helium$-$core giants by fitting their distribution with a power law plus a Gaussian, as in \citet{2006ApJ...650.1070D} (right panel of Figure~\ref{figure:2}). We used $K_{0}=-1.65$ for the intrinsic luminosity, $(J-K)_{0}=0.65$ for the intrinsic colour (infered from low-extinction \rm{2MASS} fields, van Kerkwijk personal communication) and $A_K=0.112 A_V$ \citep{1998ApJ...500..525S}. The extinction was found to range from $A_V=0.65$ to $A_V=2.64$ for distances of $1.1-4.5$\,kpc. For the 3.7\,kpc distance of \citet{2002MNRAS.337..409O}, we deduce $A_V=2.47$. Our values are larger than the ones derived by the model of \citet{2006A&A...453..635M}  (e.g. $A_V=2.52$ for 3.7\,kpc), most likely due to the different resolution of our method, but consistent with the values derived by \citet{2003A&A...409..205D}  (e.g. $A_V=2.05$ for 3.7\,kpc). 
%%%%%%%%%%%%%%%%%%%%%%%%%%%%%%%%%%%%%%%%%%%%%%%%%%%%%
\subsection{Age and Temperature}
\begin{figure}
\resizebox{8.5cm}{!}{\includegraphics{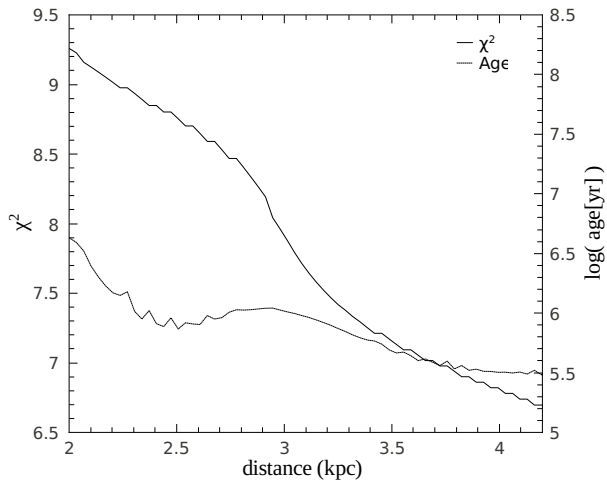}}
\caption{$\chi^{2}$ index (Eq. 1, left axis) and best-fit age (right axis) as a function of distance. The goodness of the fit continually increases with distance. The best solution is found at 4.2\,kpc where the $V$ magnitude of the WD becomes equal to the brightest value provided by the model.}
\label{figure:3}
\end{figure}

The thermodynamics of WDs are simple in nature, making the cooling rates and ages easy to calculate.
Several models exist for a wide variety of masses and compositions \citep[e.g.][]{2008AJ....135.1239H,1995PASP..107.1047B}. In the high mass domain, the colours and temperatures derived by these models are in good agreement, independently of the chemical composition, especially for ages smaller than 8\,Gyrs. Once the mass and absolute magnitudes are known, one can correlate them with a theoretical cooling track and derive the age.
In the case of the PSR\,J1141$-$6545 binary, this calculation is complicated by the uncertain distance estimate and by the fact that the measured $V-R$ color is redder than expected.
In order to find the age of the WD we used the O/Ne-core $1.06$\,M$_{\odot}$ cooling track of \citet{2008AJ....135.1239H} and searched for the best solution in the $\left\lbrace d, A_{V}, T_{\mathrm{WD}}\right\rbrace$ parameter space by minimising the quantity:
\begin{equation}
\chi^{2} = \frac{\left[ V_{0}(d,A_{V}) - V_\mathrm{WD}(T)\right] ^{2}}{\sigma_{V}^{2}} + \frac{\left[ R_{0}(d,A_{R}) - R_{\mathrm{WD}}(T)\right] ^{2}}{\sigma_{R}^{2}}
\end{equation}
with $V_{0}$ and $R_{0}$ the absolute magnitudes for a given distance and reddening; $V_{\mathrm{WD}}$ and $R_{\mathrm{WD}}$ the predicted magnitudes for a given age T; and $\sigma_{\mathrm{V,R}}$ the photometric uncertainties. We varied the distance between 2 and 4.2\,kpc with a 0.1 step size. For each distance, $A_V$ was derived from our reddening calculations. Finally, the extinction was convered using $A_R=0.819 A_V$ \citep{1998ApJ...500..525S}.
Unfortunately, our method yielded no compelling solution (Fig.~\ref{figure:3}), not only for the most reliable $1.06$\,M$_{\odot}$ track but for other \citet{2008AJ....135.1239H} and \citet{1995PASP..107.1047B} tracks of similar masses as well. In each case the minimum $\chi^{2}$ was constrained by the minimum age provided by the particular model. The impact of the results on formation scenarios of PSR\,J1141$-$6545 is discussed in the next section.
%%%%%%%%%%%%%%%%%%%%%%%%%%%%%%%%%%%%%%%%%%%%%%%%%%%%
\section{Conclusions and discussion}\label{section:4}
The results of this paper, for the first time, provide indisputable evidence for
the gravitational asymmetry of the PSR\,J1141$-$6545 binary system, i.e.\ its
composition of a strongly self-gravitating body, the pulsar ($E^{\rm
grav}/mc^2 \sim 0.2$), and a weakly self-gravitating body, the white dwarf
($E^{\rm grav}/mc^2 \sim 10^{-4}$). This is of utmost importance for testing 
alternative theories of gravity with this system, in particular tests of 
gravitational dipolar radiation. In fact, the direct observation of the white 
dwarf companion to PSR\,J1141$-$6545 substantiates limits on alternative gravity 
theories derived in the past, like in \citep{2005tmgm.meet..647E, 2008PhRvD..77l4017B}. 
Before the optical detection of the companion to PSR\,J1141$-$6545, its WD 
nature was inferred from the mass measurement, which is based on general relativity, 
and Monte-Carlo simulations of interacting binaries. These arguments are clearly 
less compeling than the evidence provided here, and become debatable when 
testing alternative theories of gravity, in particular when performing
generic tests like in the double pulsar \citep{2009CQGra..26g3001K} and the
PSR\,J1012+5307 system \citep{2009MNRAS.400..805L}.
\begin{figure}
\resizebox{8.5cm}{!}{\includegraphics{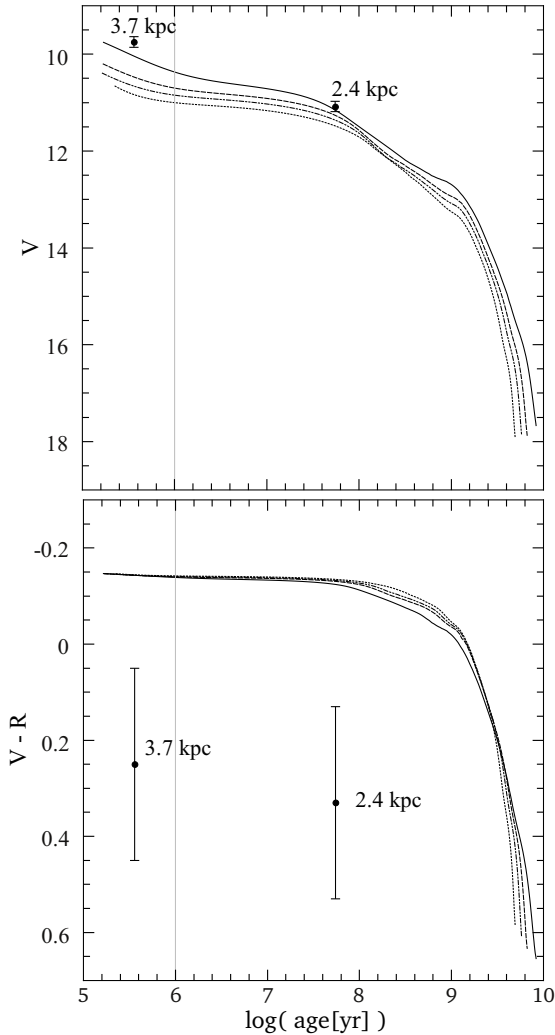}}
\caption{Cooling track of a $M=1.06$\,M$_{\odot}$ O/Ne WD (solid line) based on the work of \citet{2008AJ....135.1239H} as reflected in its $(V-R)_{0}$ colour (lower panel) and brightness (upper panel). Further WD model sequences are overplotted for comparison (masses: $1.16,1.20,1.24$\,M$_{\odot}$; dashed, dashed-dotted and dotted line respectively). The color and brightness of PSR\,J1141$-$6545 is also plotted against age, for the distance of 3.7\,kpc \citep{2002MNRAS.337..409O} and the 2.4\,kpc DM distance. The error-bars in both panels represent the $1\sigma$ uncertainties derived from monte-carlo simulations of photometric and calibration errors propagation. The grey vertical line shows the characteristic age of the pulsar.}
\label{figure:4}
\end{figure}

Concerning the cooling age of the WD, our results are discrepant with the cooling tracks of regular WDs. Therefore, we could neither confirm nor reject the proposed formation scenarios for PSR\,J1141$-$6545. In particular, as demonstrated in figure~\ref{figure:4}, if the distance to the system is 3.7\,kpc or higher \citep{2002MNRAS.337..409O} then the WD is brighter than expected and the best-fit solution to the cooling curve is found at an age smaller ($< 10^{5}$\,yrs) than the characteristic age of the pulsar. On the other hand, if the 2.4\,kpc DM distance is closer to the real one, then our results remain consistent with a WD older than the pulsar ($\sim 10^{8}$\,yrs). In both cases, the colour is redder than expected but still consistent with the cooling track within $2\sigma$. If the excessive reddening is true and not the result of an unidentified systematic error, then it can not be explained by interstellar extinction. Hence its source is most likely intrinsic to the system and possibly the consequence of an interaction between the WD and the super-nova, which polluted the atmosphere of the WD and changed its spectral fingerprint. It is worth mentioning that after performing the same kind of analysis (Section~\ref{section:3}) on the prototype eccentric WD-NS binary PSR\,B2303+46 using the $B,V$ and $R$ colours reported by \citet{1999ApJ...516L..25V}, we found that it shows similar deviations from the expected cooling track. The red color of these systems could also be explained by the presence of a cold disk around the WD which causes an excessive brightness towards the red part of the spectrum. However this scenario is less favoured because such a disk would be highly unstable given the high eccentricity of the binaries. Moreover the disk would produce orbit-correlated timing residuals that are not yet reported. Future multi-band photometric observations would be able to provide further insight into the formation history of both binaries.

\section*{Acknowledgements}
We thank Thomas Driebe for early discussions and Joris Verbiest for his recommendations on the text.
\bibliographystyle{mn2e}
\bibliography{mnras.bib}
\end{document}